\documentclass[aps,twocolumn]{revtex4}
\usepackage{epsfig}
\begin{document}


\title{
Coulomb dissociation of $^8$B and
the low-energy cross section of the $^7$Be(p,$\gamma$)$^8$B 
solar fusion reaction
}

\author{
F.~{Sch{\"u}mann}$^{1}$,
F.~{Hammache}$^{2}$,
S.~{Typel}$^{2}$,
F.~{Uhlig}$^{3}$,
K.~{S{\"u}mmerer}$^{2}$,
I.~{B{\"o}ttcher}$^{4}$,
D.~{Cortina}$^{2}$,
A.~{F{\"o}rster}$^{3}$,
M.~{Gai}$^{5}$,
H.~{Geissel}$^{2}$,
U.~{Greife}$^{6}$,
N.~{Iwasa}$^{7}$,
P.~{Koczo{\'n}}$^{2}$,
B.~{Kohlmeyer}$^{4}$,
R.~{Kulessa}$^{8}$,
H.~{Kumagai}$^{9}$,
N.~{Kurz}$^{2}$,
M.~Menzel$^{4}$,
T.~{Motobayashi}$^{9,10}$,
H.~{Oeschler}$^{3}$,
A.~{Ozawa}$^{9}$,
M.~{P{\l}osko{\'n}}$^{2,8}$,
W.~{Prokopowicz}$^{8}$,
E.~{Schwab}$^{2}$,
P.~{Senger}$^{2}$,
F.~Strieder$^{1}$,
C.~{Sturm}$^{2,3}$,
Zhi-Yu~{Sun}$^{2}$,
G.~{Sur{\'o}wka}$^{2,8}$,
A.~{Wagner}$^{11}$, and
W.~{Walu{\'s}}$^{8}$
}
\affiliation{$^1$Institut f{\"u}r Physik mit Ionenstrahlen, Ruhr-Universit{\"a}t
Bochum, D-44780 Bochum, Germany} 
\affiliation{$^2$Gesellschaft f{\"u}r Schwerionenforschung (GSI), D-64220 Darmstadt,
  Germany}
\affiliation{$^3$Technische Universit{\"a}t Darmstadt,
D-64289 Darmstadt, Germany}
\affiliation{$^4$Phillips-Universit{\"a}t Marburg,
D-35032 Marburg, Germany}
\affiliation{$^5$University of Connecticut,
Storrs, CT 06269-3046, U.S.A.}
\affiliation{$^6$Department of Physics, Colorado School of Mines, 
Golden, CO 80401, U.S.A.}
\affiliation{$^7$Tohoku University, Aoba, Sendai, Miyagi 980-08578,  
Japan}
\affiliation{$^8$Jagellonian University, PL-30-059 Krak{\'o}w, 
Poland}
\affiliation{$^9$RIKEN (Institute of Physical and Chemical Research),
Wako, Saitama 351-0198, Japan}
\affiliation{$^{10}$Rikkyo University, Toshima, Tokyo 171, 
Japan}
\affiliation{$^{11}$Forschungszentrum Rossendorf,
D-01314 Dresden, Germany}
\date{June 3, 2003}

\begin{abstract}
An exclusive measurement of the Coulomb breakup of
$^8$B into $^7$Be+p at 254 $A$ MeV allowed to study
the angular correlations of
the breakup particles.
These correlations demonstrate clearly that 
E1 multipolarity dominates
and that E2 multipolarity can be neglected. 
By using a simple single-particle model for $^8$B and 
treating the breakup in first-order perturbation theory, 
we extract a zero-energy $S$ factor of 
$S_{17}(0) = 18.6 \pm 1.2 \pm 1.0$ eV b,
where the first error is experimental and the second one reflects
the theoretical uncertainty in the extrapolation.
\end{abstract} 


\maketitle

Exciting new results \cite{Ahm01} from
the Sudbury Neutrino Observatory (SNO) have proven for the
first time that the measured high-energy neutrino flux from the
Sun agrees well with the one calculated from standard solar models
\cite{BPB01,Bru99} if non-electron flavour neutrinos are taken into
account. This again focusses attention onto the $^7$Be(p,$\gamma$)$^8$B
reaction which provides almost exclusively the high-energy neutrinos
measured in the SNO experiment. Their flux depends linearily
on the $^7$Be(p,$\gamma$)$^8$B cross section
at solar energies. Very recently, the latter has been redetermined
by new high-precision direct measurements 
\cite{Ham01,Str01,Jun02,Bab03} and extrapolated to zero energy with
the help of a theoretical model \cite{Des94}. The resulting
zero-energy astrophysical $S$ factors, $S_{17}(0)$,
however, do not always agree within their quoted
errors:
Hammache {\it et al.} \cite{Ham01} found $S_{17}(0) = 18.8 \pm 1.7$ eV b,
in agreement with other direct-capture data \cite{Fil83,Vau70,Str01}. 
In contrast, Junghans {\it et al.} \cite{Jun02} report a
considerably larger value, $S_{17}(0) = 22.3 \pm 0.7 \pm 0.5$ eV b.
The very recent result of Baby {\it et al.} \cite{Bab03} also
favours a rather large value of $S_{17}(0) = 21.2 \pm 0.7$ eV b.

In view of their importance for astro- and elementary-particle
physics, these conflicting results should be 
verified and cross-checked by other, indirect
measurements that have different systematic errors. One possibility
is Coulomb dissociation (CD) of $^8$B in the electromagnetic field
of a high-$Z$ nucleus. 
Such measurements have been performed at low \cite{Kol01}, intermediate
\cite{Kik98,Dav01}, and high energies \cite{Iwa99}.
Alternatively, $S_{17}$(0) can
also be calculated from asymptotic normalization coefficients
(ANC) which in turn are determined in low-energy proton-transfer
or in proton-removal
reactions \cite{Azh01,Tra01,Cor03}. 

In the present Letter, we focus on a crucial question that must be
answered if one wants to use the CD method 
to derive a precise value for $S_{17}$(0). 
The astrophysical $S$ factors of the $^7$Be(p,$\gamma$)
reaction can only be calculated reliably from the energy-differential
CD cross sections if the electromagnetic multipole
components relevant for direct capture and the time-reversed
process have the same strength. In low-energy proton capture
the E1 contribution by far dominates the cross section.
While E1 is the dominant multipolarity
also in CD, one can show easily that the
equivalent photon field emitted from a high-$Z$ target nucleus
contains a strong E2 component. This is particularly true for
CD at low energies. At higher energies
(see Ref.~\cite{Iwa99}) 
the relative amount of E2 multipolarity is expected to be reduced,
but may still be substantial enough to affect the final result.
To remove this ambiguity,
it is indispensable to either determine the E1/E2 ratio in
CD experimentally, or to extract $S_{17}$  with such cuts
that any E2 contribution is negligible. 

Experimental limits for a possible E2 contribution
were extracted in the work of Kikuchi {\it et al.}~\cite{Kik98}
and Iwasa {\it et al.}~\cite{Iwa99}. Both papers found negligible
E2 contributions.
Recently, Davids {\it et al.} have reported
positive experimental evidence for a finite E2 contribution in CD
of $^8$B, mainly from the analysis of {\em inclusive} longitudinal 
momentum ($p_{||}$) spectra of $^7$Be
fragments measured at 44 and 81 $A$ MeV \cite{Dav01}.
The asymmetries in the $p_{||}$ spectra
were interpreted to be due to E1-E2 interference
in terms of first-order perturbation-theory~\cite{Esb96}.

In order to resolve these discrepancies,
we decided to perform an {\em exclusive} CD experiment at 
high energy (254 $A$ MeV) 
at the kaon spectrometer KaoS at GSI~\cite{Sen93}
with the aim to measure quantities that should be sensitive
to contributions of E2 multipolarity, namely the angular correlations of
the $^8$B-breakup particles, proton and $^7$Be. Experimentally, this
requires high-resolution measurements of the positions and angles of
the incident $^8$B beam
as well as those of the breakup fragments.
The $^8$B secondary beam was produced at the SIS/FRS radioactive
beam facility at GSI~\cite{Gei92} by fragmenting 
a 350 $A$ MeV $^{12}$C beam in a
8 g/cm$^2$ Be target and separating it from contaminant ions in
a 1.4 g/cm$^2$ wedge-shaped Al degrader 
placed in the FRS intermediate focal plane.
Typical $^8$B beam intensities in front of KaoS were $5 \times 10^4$ per
4 sec spill; the only contaminant consisted of about 20\% $^7$Be
ions which could be identified event by event with the help of a
time-of-flight measurement.

Positions and angles of the secondary beam incident
on the Pb breakup target were measured with the help of two parallel-plate
avalanche counters (PPAC) located at 308.5 cm and 71 cm upstream from the
target, respectively. The detectors, which were designed at RIKEN
\cite{Kum01}, had areas of $10 \times 10\,{\rm cm}^2$ 
and allowed to track the incident $^8$B beam 
with about 90\% efficiency and with position and angular
resolutions of 1.3 mm and 1 mrad, respectively. Downstream from
the Pb target (which consisted of 50 mg/cm$^2$ $^{208}$Pb
enriched to 99.0$\pm$0.1\%),
the angles and positions as well as the energy losses
of the outgoing particles were measured with two pairs of Si
strip detectors (300 $\mu$m thick, 100 $\mu$m pitch) located at distances
of about 14 cm and 31 cm. Proton and $^7$Be momenta were
analyzed with the KaoS spectrometer 
which was set up almost identical
\begin{figure}[bt]
\epsfig{file=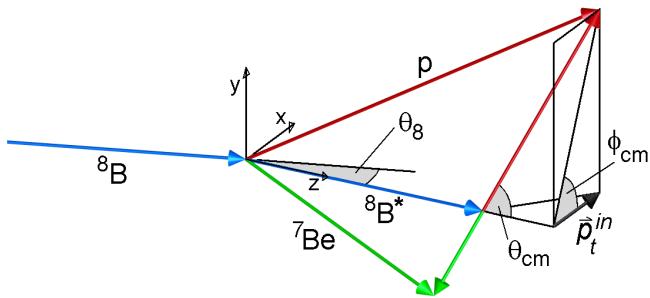,height=4cm}
\caption{Vector diagram showing the definitions of the angles 
$\theta_{cm}$ and $\phi_{cm}$ as well as the proton in-plane transverse
momentum, $p_t^{in}$, 
in the frame of the $^{8}$B$^*$ system.}
\label{vector_diagram}
\end{figure}
to our previous experiment \cite{Iwa99}, except for a newly constructed
plastic-scintillator wall near the KaoS focal plane with 30 elements
(each 7 cm wide and 2 cm thick) used for trigger purposes. 

The coincident p and $^7$Be signals
resulting from breakup in the $^{208}$Pb target were identified by
reconstructing their vertex at the target, this removed all breakup
events in layers of matter other than the target.
The measured momentum vectors of the outgoing p and $^7$Be particles
allowed to construct the invariant-mass spectrum of
the excited $^8$B$^*$ system prior
to breakup.
\begin{figure}[t]
\vspace*{-2mm}
\epsfig{file=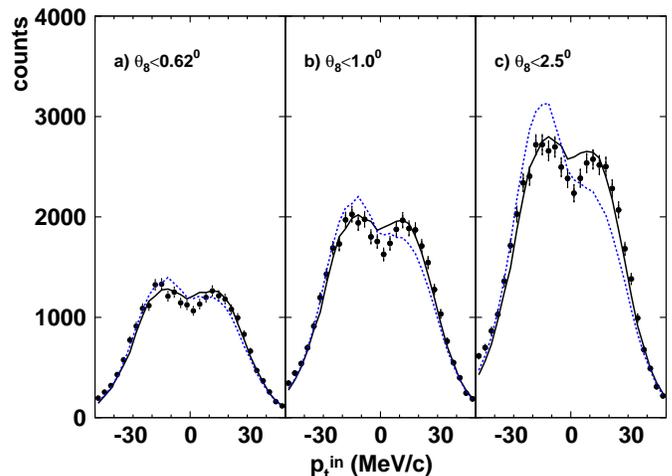,width=9cm}
\vspace*{-1.1cm}
\caption{
In-plane transverse momenta, $p_t^{in}$, of the break-up protons
for three different cuts in $\theta_8$.  
The theoretical curves (full lines: E1 multipolarity, dashed
lines: E1+E2 multipolarity) have been calculated 
in first-order perturbation theory.
They were normalized individually to the
data points in each frame.
}
\label{ptin}
\end{figure}
Fig.~\ref{vector_diagram} shows the coordinate systems used.  
The angle $\theta_8$ is the laboratory scattering angle of
$^8$B$^*$ relative to the incoming $^8$B beam.
The polar angles, $\theta_{cm}$, and the azimuthal
angles, $\phi_{cm}$, of the breakup protons are measured 
in the rest frame of the $^8$B* system, as shown
in Fig.~\ref{vector_diagram}.
In the same way, one can calculate e.g. the transverse proton momentum vector
in the reaction plane ($p_t^{in}$).

In the following we will present some angular distributions of the
emitted proton in the frame of the $^8$B$^{*}$ system that can be 
shown to be sensitive to an E2 amplitude in CD.
To interprete the measured distributions we need guidance by a theoretical
model. To this end,
we have performed standard first-order perturbation-theory (PT)
calculations of the CD process in the semi-classical
approach~\cite{Ber94,Typ02}, using a
simple Woods-Saxon potential model for $^8$B.
The potential depth for the ground state of $^8$B
was adjusted to match the proton binding energy;
the potential depths of the scattering
states were fitted to
the scattering lengths of the $^7$Li+n mirror system~\cite{Koe83}.
We used a radius parameter of
$r_0=1.25$ fm and a diffuseness of $a=0.65$ fm.
For channel spin $I=2$ (the dominant contribution) 
we obtained a potential
depth of $V_2=52.60$ MeV. 
The resulting scattering length for this channel of
$a_{02}^{theo} = -8$ fm agrees well with the recently measured value
of $a_{02}^{exp} = -7 \pm 3$ fm (Angulo {\it et al.}~\cite{Ang03}).

To take into account absorption
due to nuclear overlap in CD, we have introduced a diffuse 
absorptive nuclear potential
with a depth of 20 MeV and a radius of 9.91 fm, i.e. the sum of
the projectile and target radii. This choice
reproduces well the integral $\theta_8$ angular distribution.
Technically, the results of the PT calculations were returned as
a statistically-distributed ensemble of $500\,000$ CD-``events''
that were analyzed in the same way 
as the experimental data, thus imposing the experimental cuts.  

We first present in Fig.~\ref{ptin} the distribution
of $p_t^{in}$ for three different upper limits in $\theta_8$,
0.62$^{\circ}$, 1.0$^{\circ}$, and 2.5$^{\circ}$. 
In classical Rutherford scattering, this corresponds to impact parameters
of 30 fm, 18.5 fm, and 7 fm, respectively.
Relative energies between p and $^7$Be
up to 1.5 MeV were selected.
The experimental data 
for all three $\theta_8$-cuts can be reproduced well
by a PT calculation
that includes only E1 multipolarity
(full histograms in Fig.~\ref{ptin}, the
theoretical curves were normalized individually to the data points).
If E1-plus-E2 multipolarity is used in the PT calculation,
the different impact-parameter dependences of E1 and E2 multipolarity
lead to markedly different shapes for the different $\theta_8$-cuts
(dashed histograms in Fig.~\ref{ptin}). 
The latter distributions are, however, 
in clear disagreement with our data points.

\begin{figure}[bt]
\vspace{-0.8cm}
\epsfig{file=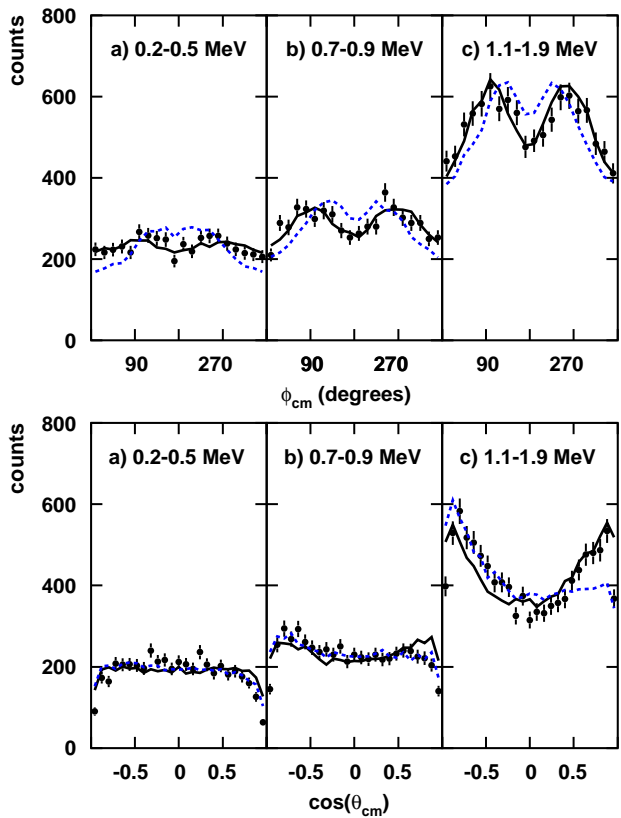,width=9cm}
\vspace*{-1.0cm}
\caption{
Top: Experimental distributions of the proton azimuthal 
angular ($\phi_{cm}$) distributions
for three different bins of the p-$^7$Be relative energy, $E_{rel}$.
The full histograms denote a first-order perturbation-theory calculation
for E1 multipolarity, the dashed ones for E1+E2.
All theoretical curves were individually normalized to the
data points in each frame.
Bottom: the same for the polar breakup angles, $\theta_{cm}$.
}
\label{angular_distributions}
\end{figure}

Fig.~\ref{angular_distributions} depicts the experimental 
$\phi_{cm}$ and $\theta_{cm}$ distributions for three
different $E_{rel}$ bins,
as indicated in the figure. A ``safe'' $\theta_8$ 
limit of 1$^{\circ}$ was chosen.
As expected, these distributions are mostly isotropic at low $E_{rel}$
(indicative of $s$-waves)
and become increasingly anisotropic for larger values (contributions
from $d$-waves).
For the $\phi_{cm}$ distributions, which are most sensistive to E2 admixtures,
the calculations for pure E1 multipolarity clearly fit best;
inclusion of an E2 component
shifts the maxima away from 90$^{\circ}$ and 270$^{\circ}$
with increasing $E_{rel}$,
while at the same time the anisotropy is reduced.
Similar conclusions can be drawn
from the bottom part of Fig.~\ref{angular_distributions}, where
the proton polar angular ($\theta_{cm}$) distributions are shown.
The low-$E_{rel}$ bins show little sensitivity to E2 multipolarity,
whereas inclusion of E2 leads to a marked 
discrepancy near $\cos(\theta_{cm}) = 1$
for the highest $E_{rel}$ bin.
More detailed calculations show that at most E2 amplitudes of
$\lesssim$0.3 times the theoretical one from 
our simple model are simultaneously 
compatible with all our measured observables. 
Since this would correspond to E2-contributions
to the cross sections of less than 1\%, much less than the errors of the
data points, we neglect the effect of E2 multipolarity. 
This is in line with conclusions drawn by 
Kikuchi {\it et al.} \cite{Kik98}
and by Iwasa {\it et al.} \cite{Iwa99}
from their respective $\theta_8$ distributions (which are, however,
less sensitive to a small E2 component than the present angular
correlations). 
Our findings contradict the conclusions of Davids {\it et al.} \cite{Dav01}
that a substantial E2 cross section has to be subtracted
from the total measured CD cross section.

Our results allow to interpret the relative-energy distributions
of the breakup particles
in an easy way. In the following, we have 
restricted the angles $\theta_8$ 
to values below 0.62$^{\circ}$ to ensure both dominance
of CD and reduction of the effect of any possible
E2 contribution. 
The data are compared to a simulation with GEANT that
includes two electromagnetic multipole components:
a resonant M1 contribution located at $E_{rel}$=0.63 MeV with resonance
parameters taken from
Filippone {\it et al.}~\cite{Fil83},
and the non-resonant E1 contribution from our theoretical
model as described above. The latter was scaled by a
normalization factor of 0.79.
Note that we have added to the GEANT simulation a contribution
that feeds the first excited state in $^7$Be at 429 keV using the
measurements of Kikuchi {\it et al.} \cite{Kik98}.  
\begin{figure}[bt]
\vspace*{-0.3cm}
\epsfig{file=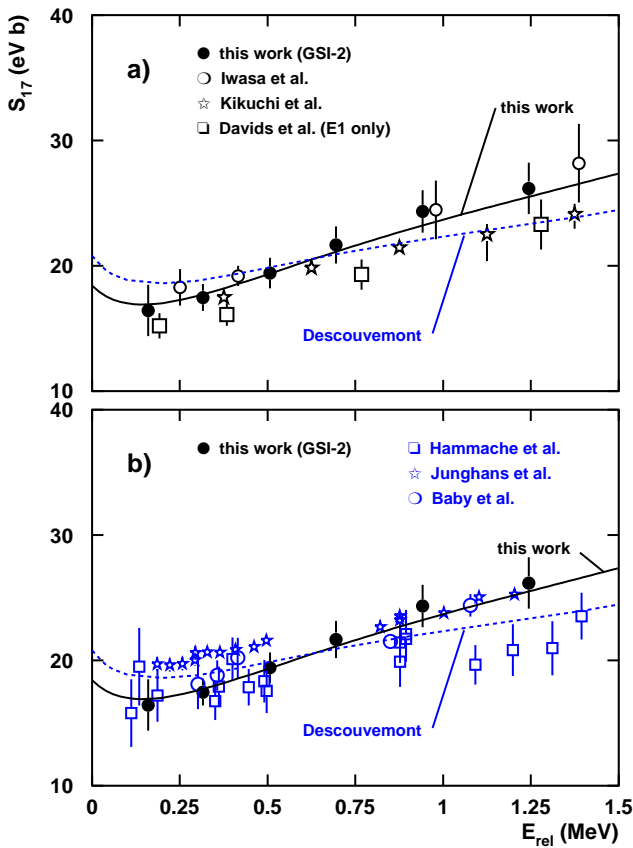,width=9.0cm}
\vspace*{-11mm}
\caption{
a) Comparison between $S_{17}$ values from
Coulomb-dissociation experiments. 
The full (open) circles indicate the present
(previous) GSI CD experiment. 
Open stars depict Ref.~\protect\cite{Kik98}, 
open squares Ref.~\protect\cite{Dav01} (E2 
contribution subtracted).
The theoretical curves are described in the text.\\
b) $S_{17}$ from this work in comparison with 
the (p,$\gamma$) experiments
of Ref.~\protect\cite{Ham01} (squares), 
Ref.~\cite{Jun02} (stars), and Ref.~\cite{Bab03} (open circles).
The latter data were corrected for the contribution of the M1 resonance
by the authors.
}
\label{S_17}
\end{figure}
Subtracting the small M1 contribution (that affects only a narrow $E_{rel}$
region around the resonance), the remaining $d\sigma/dE_{rel}$ distribution can
be converted to the E1 astrophysical $S$ factor
$S_{17}(E_{rel})$.

The resulting $S_{17}$ factors (averaged over $E_{rel}$ bins 
0.2 to 0.3 MeV wide)
are visualized in Fig.~\ref{S_17}.
The error bars do not include a common systematic error of 5.6\%.
The top panel (a) compares our results to those of other CD experiments
\cite{Kik98,Iwa99,Dav01} (the data of Ref.\cite{Dav01} 
represent their E1-$S_{17}$ factors after subtraction of the E2-contribution).
At low $E_{rel}$, the CD $S$ factors are in good agreement,
though the Davids {\it et al.}~\cite{Dav01} data are systematically lower.
The bottom panel (b) compares our data to those of the 
$^7$Be(p,$\gamma$)$^8$B measurements where the authors have subtracted the
contribution from the M1 resonance (Refs.~\cite{Ham01,Jun02,Bab03}).
At low energies the (p,$\gamma$) data of Refs.~\cite{Ham01,Bab03}
and ours are in good agreement,
whereas the Seattle data~\cite{Jun02} deviate considerably.
The opposite behaviour is noted above the M1 resonance:
our data and those of Refs.~\cite{Jun02,Bab03} match excellently,
whereas the other (p,$\gamma$) experiments~\cite{Ham01,Str01,Vau70,Fil83}
consistently report lower values.
We want to emphasize the remarkably 
good agreement of our CD data up to 1.1 MeV with the most recent
direct-proton-capture experiment
where an ion-implanted $^7$Be target was used~\cite{Bab03}.

To extrapolate to zero energy,
all recent (p,$\gamma$) experiments have chosen the
cluster model of Descouvemont and Baye~\cite{Des94}.
When we fit our data points up to $E_{rel}$ = 1.5 MeV
to this model and add in quadrature a common systematic error of 5.6\%,
we obtain $S_{17}(0) = 20.8  \pm 1.3 $ eV b
(dashed lines in Fig.~\ref{S_17}). Restricting the fit to
energies below 0.6 MeV, where the model-dependence
has been shown to be weaker~\cite{Jen98},
$S_{17}(0) = 19.6  \pm 1.4$ eV b is obtained.  
Our potential model, however, reproduces the data
over the entire energy range up to 1.5 MeV, 
yielding $S_{17}(0) = 18.6  \pm 1.2$ eV b
(full lines in Fig.~\ref{S_17}). It is interesting to note that
a fit of the Baby {\it et al.} (p,$\gamma$) data to our model
yields practically the same result,
$S_{17}(0)=18.1 \pm 0.3$ eV b.
Clearly, still more high-precision experimental data are needed 
to resolve the discrepancies between the experimental data sets and
to pin down the correct theoretical extrapolation of the measured data
to solar energy.
In the meantime, an additional ``extrapolation error'' of
$\pm$~1.0 eV b 
seems appropriate.

We conclude that Coulomb dissociation has been proven to be a
valuable method to provide a rather precise value for the 
low-energy $^7$Be(p,$\gamma$) cross section.
Since in CD all energy bins are measured simultaneously,
CD provides a reliable 
measurement of the {\it shape} of the $S_{17}$ distribution.
By setting tight constraints to the scattering angle
$\theta_8$ and analyzing p-$^7$Be angular correlations, 
a significant contribution
from E2 multipolarity can be excluded. 
Small modifications of the Woods-Saxon 
potential parameters allow to reproduce the data 
in first-order perturbation theory
with remarkable accuracy
up to about $E_{rel} = 1.5$ MeV.

The authors wish to thank K.-H.~Behr, K.~Burkard, and A.~Br{\"u}nle
for technical assistance. Vivid discussions with B.~Davids,
P.~Descouvemont, M.~Hass, and A.~Junghans are gratefully acknowledged.

\end{document}